\documentclass[11pt]{article}

\usepackage[margin=1in]{geometry}
\usepackage{setspace}
\usepackage[utf8]{inputenc}
\usepackage[T1]{fontenc}
\usepackage[english]{babel}
\usepackage{lmodern, csquotes, eurosym}
\usepackage[style=authoryear,texencoding=utf8,backend=biber]{biblatex}
\usepackage[colorlinks=true]{hyperref}
\usepackage{tikz}
\usepackage{caption}
\usepackage{subcaption}
\usepackage{mathtools}
\usepackage{tabularray}
\usepackage{tabularx,calc,array}
\usepackage{amssymb}
\usepackage{float}
\usepackage{amsfonts}
\usepackage{geometry}
\usepackage{graphicx}
\usepackage{booktabs}
\usepackage{pdflscape}
\newtheorem{theorem}{Theorem}[section]
\newtheorem{mf}[theorem]{Main Finding}

\renewbibmacro{in:}{%
  \ifentrytype{article}
    {}
    {\bibstring{in}%
     \printunit{\intitlepunct}}}

\title{Perceptions and Detection of AI Use in \\ Manuscript Preparation for Academic Journals
\thanks{We thank many members of the UCSB Economics Department, the Decision Theory Forum, and the Economic Science Association, especially Monica Capra, for completing our survey and for very helpful suggestions. We also thank Ambreen Chaudhri at Kellogg Research Support for her help in building our manuscript database,  and to Ruixi (Raye) Zhu, Marisa Weng, and Bonny Wang for research assistance.}
}
\author{Nir Chemaya\thanks{University of California, Santa Barbara, \href{mailto:nir@ucsb.edu}{nir@ucsb.edu}}\: and Daniel Martin\thanks{University of California, Santa Barbara and Kellogg School of Management, Northwestern University, \\ \href{mailto:danielmartin@ucsb.edu}{danielmartin@ucsb.edu}.}
}

\date{November 24, 2023}

\begin{document}

\maketitle

\begin{abstract}
The emergent abilities of Large Language Models (LLMs), which power tools like ChatGPT and Bard, have produced both excitement and worry about how AI will impact academic writing. In response to rising concerns about AI use, authors of academic publications may decide to voluntarily disclose any AI tools they use to revise their manuscripts, and journals and conferences could begin mandating disclosure and/or turn to using detection services, as many teachers have done with student writing in class settings. Given these looming possibilities, we investigate whether academics view it as necessary to report AI use in manuscript preparation and how detectors react to the use of AI in academic writing.
\end{abstract}

\newpage 

\section{Introduction}

There is both excitement and concern about the impact that artificial intelligence (AI) could have on our world. In academia, this has inspired an explosion of research around AI. Some researchers are leveraging AI to help in answering research questions; for example, as a tool for performing statistical or textual analyses (e.g., \cite{mullainathan2017machine}, \cite{athey2019machine}, \cite{fudenberg2019predicting}, \cite{farrell2020deep}, \cite{rambachan2021identifying},   \cite{björkegren2022machine}, \cite{capra2023sound}, \cite{franchi2023detecting}, \cite{salah2023may}), or for the design or implementation of experiments (e.g., \cite{beck2020artificial}, \cite{charness2023generation}, \cite{horton2023large}).  Others have focused more on the interaction of humans with AI systems (e.g., \cite{dietvorst2015algorithm}, \cite{deza2019assessment}, \cite{gajos2022people},  \cite{steyvers2022bayesian}, \cite{sundar2022rethinking}, \cite{tejeda2022ai}, \cite{wang2023effects}, \cite{yang2023gender}) and the larger societal ramifications of AI (e.g., \cite{agrawal2019economics}, 
\cite{lambrecht2019algorithmic}, \cite{obermeyer2019dissecting}, \cite{yangdoes}, \cite{chien2020artificial}, \cite{rolf2020balancing}, \cite{zuiderwijk2021implications}, 
\cite{pallathadka2023applications}, \cite{ray2023chatgpt}, \cite{singh2023chatgpt}).

In terms of societal ramifications, one topic of interest is how AI will impact academia itself, especially given the emergent abilities of Large Language Models (LLM), which power tools like ChatGPT and Bard. Most of this research has centered on student use of AI to complete course or degree requirements (e.g., \cite{cowen2023learn}, \cite{daun2023chatgpt}, 
\cite{fyfe2023cheat}, \cite{ibrahim2023rethinking}, \cite{jungherr2023using}, 
\cite{malik2023exploring}, \cite{schmohl2020artificial}, \cite{shahriar2023let}). However, there has been less focus on the impact that AI is having and might have on manuscript preparation for academic journals, even though the impact this could have on science, and the propagation of scientific results, is potentially large.

An exception is \textcite{korinek2023language}, who documents several use cases for LLMs for researchers, including academic writing. He notes, ``LLMs can edit text for grammatical or spelling mistakes, style, clarity, or simplicity.'' As a result, this class of tools ``allows researchers to concentrate their energy on the ideas in their text as opposed to the writing process itself.'' In particular, ``this set of capabilities are perhaps most useful for non-native speakers who want to improve their writing.'' The potential impact of these tools is validated by the emergence of companies that help academics leverage LLM in their writing. Some of those initiatives are even run by professors from academia, such as online workshops that teach researchers how to use ChatGPT for academic publishing.

However, an emerging set of papers (e.g., \cite{altmae2023artificial}, \cite{thorp2023chatgpt},  
\cite{shahriar2023let} and \cite{hill2023chat}) detail some practical and ethical concerns with the use of these tools in preparing manuscripts for academic journals. For example, AI can generate a text with mistakes, including incorrect math, reasoning, logic, factual information, and citations (even producing references to scientific papers that do not exist). There are many well-documented examples where LLM's ``hallucinate'' and provide completely fictitious information. On top of this, these tools may also produce text that is biased against particular groups. These issues are exacerbated by the ``black box'' nature of LLM suggestions, meaning that we lack an understanding of how they work  (e.g., \cite{bommasani2021opportunities}). Given these issues, liabilities arise when authors submit papers without fully vetting the text generated by LLMs. In addition, given that they are trained on a corpus of other writing and not the author's own writing, using the output of tools like ChatGPT and Bard without proper attribution could be considered plagiarism.  

Because of these concerns, authors may voluntarily choose to disclose whether they have used LLMs in preparing their manuscripts, as suggested by \textcite{bom2023exploring}. However, given well-documented failures of disclosure in the field and lab (e.g., \cite{dranove2010quality}, \cite{jin2015no}), journals, conferences, and associations may respond by mandating disclosure through reporting requirements, as is already done for conflict-of-interest issues. For example, Elsevier limits the use of AI by authors ``only to improve the language and readability of their paper'' and requires ``the appropriate disclosure... at the bottom of the paper in a separate section before the list of references.'' See Table~\ref{tab:Publishers} in Appendix B for a list of major publishers who have voluntary and mandatory disclosure policies for AI use. To enforce these reporting requirements, journals, conferences, and associations might turn to using detection services, as many teachers have done with student writing in class settings. The use of LLMs in writing is potentially detectable because as \textcite{korinek2023language} notes, ``Some observe that [LLM writing] is, naturally, a bit sterile and lacks the idiosyncrasies and elements of surprise that characterize human writing -- a feature that detectors of LLM-written text zero in on.''

Given these looming possibilities, we investigate whether academics view  reporting AI use in manuscript preparation as necessary and how detectors react to the use of AI in manuscript preparation. We first conducted a survey of academics in which we elicited perceptions about reporting assistance in manuscript preparation. We then used GPT-3.5 to revise abstracts from the past 10 years of \textit{Management Science} by asking the software program to either fix grammar or rewrite text. Finally, we ran the original abstracts and revised abstracts through a leading paid AI detection service.

\begin{figure*}
\centering
    \begin{subfigure}[b]{.435\textwidth}
    \centering
    \includegraphics[width=1\linewidth]{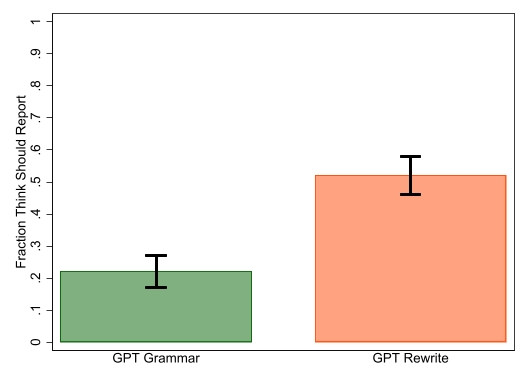}
    \caption{Fraction of survey respondents indicating that ChatGPT use in fixing grammar or rewriting text should be reported, with 95\% confidence intervals.}
    \end{subfigure}
    \hfill \begin{subfigure}[b]{.435\textwidth}
    \centering
    \includegraphics[width=1\linewidth]{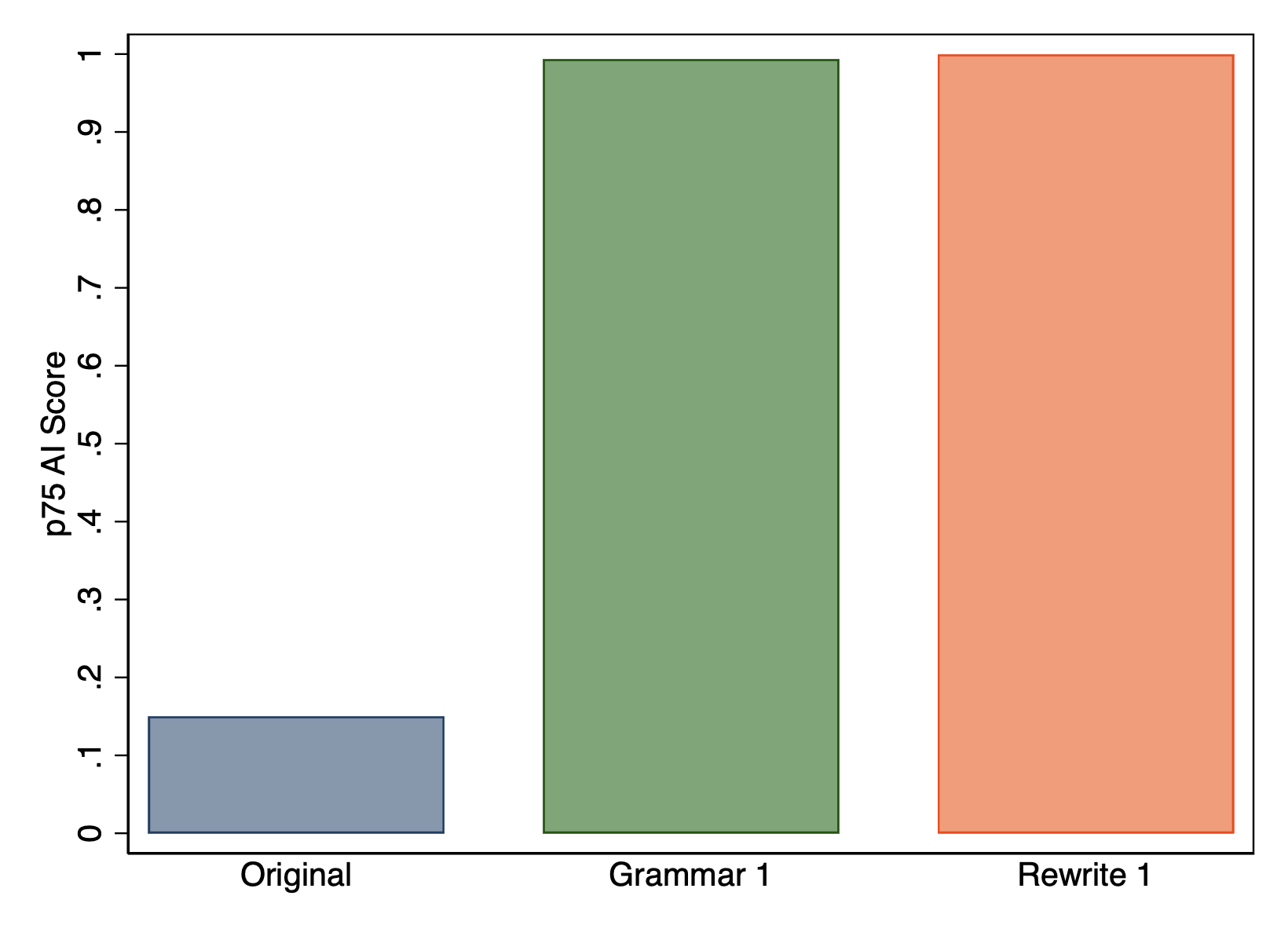}
    \caption{75th percentile of AI score for original abstracts and the versions that GPT-3.5 was used to either fix grammar or rewrite text.}
    \end{subfigure}
    \caption{Reporting views (a) vs. detection results (b).}
    \label{fig:main1}
\end{figure*}

We have three main findings. 
First, as shown partially in Figure~\ref{fig:main1}, the academics we surveyed were less likely to think that using AI to fix the grammar in manuscripts should be reported than using AI to rewrite manuscripts, but detection software did not always draw this distinction, as abstracts for which GPT-3.5 was used to fix grammar were often flagged as having a high chance of being written by AI. 
Second, we found little difference in preferences for reporting ChatGPT and research assistant (RA) help, but significant differences in reporting preferences between these sources of assistance and paid proofreading and other AI assistant tools (Grammarly and Word).
Third, we found disagreements among the academics we surveyed on whether using ChatGPT to rewrite text needs should be reported, and differences were related to perceptions of ethics, academic role, and English language background.

The rest of the paper is structured as follows. Section 2 introduces our survey design and our method for testing ChatGPT use with AI detection algorithms. Section 3 shows our main results. Finally, section 4 discusses some limitations of our work, open questions arising from our results, and suggestions for future research. 

\section{Methods}

In order to answer our main research questions, we split our design into two parts. The first part was a survey of the academic community that captured academics' views on reporting AI use in manuscript preparation. The second part was to test an AI detector's reaction to AI use in manuscript preparation. Our primary focus was on using AI to fix grammar and rewrite text, which we consider to be important intermediate cases between not using AI at all and using AI to do all of the writing based on limited inputs (e.g., only a title and/or a set of results). We asked academics for their perceptions about these two kinds of AI usage in manuscript preparation and investigated how the detector responded to these different kinds of usage. All of this will be explained in more detail in the following subsections. 

\subsection{Survey Design}

Beginning August 22nd 2023 and running until September 20th 2023, we conducted a brief unpaid survey about perceptions of AI use in manuscript preparation. We sent this survey to a convenience sample of academics; specifically, we sent it to three listservs: the UCSB Economic Department listserv (connecting graduate students, professors, and postdocs at the University of California, Santa Barbara); the Economics Science Association (ESA) announcement listserv (ESA is the leading organization for experimental economics); and the Decision Theory (DT) Forum listserv (the DT Forum is a listserv of academics working on decision theory that is run by Itzhak Gilboa). We sent the survey with a few days delay between each group, which allows for a rough measure of the respondents from each source. We had a total of 271 respondents complete the survey: 38 from UCSB, 199 from UCSB and the ESA community, and 34 from UCSB, the ESA community, and the DT Forum (this number does not include 20 individuals who did not specify an academic role). The median time to answer the survey was 1.21 minutes, and survey respondents provided informed consent (as shown in Appendix A).

In this survey, we focused on two main aspects related to perceptions of AI use in manuscript preparation. The first aspect is whether authors should acknowledge using ChatGPT to fix grammar or to rewrite text. We asked this over two separate questions, as shown in Figure~\ref{fig:part1} in Appendix A. These two uses of ChatGPT reflect to many of the use cases of LLMs suggested in articles, websites, and online courses. The second aspect is whether using ChatGPT to modify an academic manuscript is unethical. Again, we split this into two questions -- one for grammar and one for rewrite -- to allow researchers to address whether ethics also depends on usage. Because we suspected that there might be differences in these perceptions and an academic's role and language background, we asked respondents if they were a native speaker of English and their current role (postdoc, student, untenured professor, tenured professor, and/or other). Of the 271 respondents who completed our survey, 83 reported being a native English speaker, and we categorized 67 respondents as students, 32 as postdocs, 59 as untenured professors, and 113 as tenured professors. If someone reported multiple roles, we took the ``highest'' report, ranked in this order.

In addition to these questions, we added follow-up questions on a second page of the survey about whether authors should acknowledge other writing services and tools such as Word, Grammarly, proofreading, and the work of research assistants (see Figure~\ref{fig:part2} in Appendix A). We asked this question only after the ChatGPT questions and on a new page because our main focus was the perception of AI (e.g., ChatGPT) in academic writing, and because we wanted to avoid these follow-up questions influencing the answers given about ChatGPT. In addition, we did not randomize the order between ChatGPT and other services because the convenience sample already knew this was a survey about ChatGPT, so that would have been in their mind already when completing the survey. However, by asking questions about ChatGPT first, we might have reduced differences in reported perceptions with other writing services, as people might feel the need to report consistently for a given use.

To make the questions about other sources comparable the questions about ChatGPT, we used the same question format and split the comparison into two groups: fixing grammar and rewriting text. The tools we included for fixing grammar included Word, Grammarly, and RA help. Importantly, we wrote explicitly in the question that these tools were to be used for fixing grammar. For rewriting text, we used proofreading and RA help. Once more, we explicitly asked whether authors should acknowledge using these tools and services for rewriting text for an academic journal.

\subsection{Detection Design}

Next, we tested how an AI detector would react to using ChatGPT to fix grammar in academic text and to rewrite academic text. We first collected titles and abstracts from 2,716 papers published in the journal \textit{Management Science} from January 2013 to September 2023. We excluded articles with the following words in their titles, as they did not appear to be original articles: ``Erratum,'' ``Comment on,'' ``Management Science,'' ``Reviewers and Guest Associate Editors,'' and ``Reviewers and Guest Editors.'' We intentionally included papers that were published before the launch of ChatGPT in November 30, 2022 in order to have source text that was plausibly unimpacted by LLM use.

We revised these abstracts using a variety of prompts, as seen in Table~\ref{tab:table1} below. Rather than selecting the prompts ourselves, we wanted to use an external source. We decided to use prompts from the online GitHub page \textit{ChatGPT Prompts for Academic Writing}, which advises researchers on how to use ChatGPT.  We chose this source because it was the first link returned from a Google search of ``ChatGPT Prompts for Academic Writing.''

\begin{table}[H]
    \begin{tblr}{|c|X[j,valign=m]|}
        \hline
        & Prompt \\
        \hline 
        Grammar 1 & Correct the grammar. Give a version in one paragraph based on this paragraph: ``[PARAGRAPH]''   \\
        \hline
        Rewrite 1 &   Rewrite this paragraph in an academic language. Give a version in one paragraph based on this paragraph: ``[PARAGRAPH]''  \\
                \hline
        Grammar 2 & Act as a language expert, proofread my paper on “[TITLE]” while putting a focus on grammar and punctuation. Give a version in one paragraph based on this paragraph: ``[PARAGRAPH]''\\
        \hline
         Rewrite 2 & Improve the clarity and coherence of my writing. Give a version in one paragraph based on this paragraph: ``[PARAGRAPH]''\\
        \hline 
        Grammar 1b & Correct the grammar: ``[PARAGRAPH]''   \\
        \hline
        Rewrite 1b &   Rewrite this paragraph in an academic language: ``[PARAGRAPH]''  \\
             \hline
    \end{tblr}
\caption{Prompts used to revise abstracts with GPT-3.5.}
\label{tab:table1}
\end{table}

Our baselines prompts were ``Grammar 1'' and ``Rewrite 1,'' and for robustness we also considered two functionally related prompts, ``Grammar 2'' and ``Rewrite 2.'' We added a restriction at the end of these prompts to ``Give a version in one paragraph based on this paragraph'' in order to meet the requirement that the abstract be only one paragraph in length. To consider the robustness of our results to this additional language, we also ran versions of our baseline prompts that did not include it: ``Grammar 1b'' and ``Rewrite 1b.''

It has been shown that there are ways to fool certain detectors, such as by adding strange characters to the text, and this is a constantly evolving game of cat-and-mouse as detectors evolve. Although this could be of interest for researchers, we do not study this particular phenomenon. Instead, it is our goal to see whether an academic who uses ChatGPT for widely-proposed purposes -- without further edits or detection avoidance strategies -- would have their writing be flagged as AI generated.

To revise these abstracts at scale, we leveraged the GPT API with settings that would produce output that closely mirrors the output of a researcher using ChatGPT to revise academic writing. The specific model we used was gpt-3.5-turbo-0613 (GPT-3.5 Turbo released in June 13 2023), and we kept the default settings for temperature, top p, frequency penalty, and presence penalty. We also used the default \textit{system} prompt, ``You are a helpful assistant,'' which is a background prompt that can be changed in the API but not in ChatGPT itself. We did a fresh call each time we used a prompt to avoid learning and history effects.

Finally, we used a leading paid service (Originality.ai) to see how AI detection algorithms might react to this use of LLMs. We first evaluated the original \textit{Management Science} abstracts, and then the abstracts revised by GPT-3.5 based on all of the prompts. Originality.ai provides an ``AI score,'' which is a value between 0\% and 100\% that is interpreted as the likelihood that AI wrote the text being evaluated. A high score means a high likelihood that AI generated the text. Specifically, the company states: ``If an article has an AI score of 5\%... there is a 95\% chance that the article was human-generated (NOT that 5\% of the article is AI generated).'' \textcite{akram2023empirical} studied a number of popular detection tools and found that Originality.ai had the highest accuracy rate (97\%).

\section{Results}

\subsection{Reporting Views and Detection Evaluations}

Figure~\ref{fig:main1} presents the fraction of survey respondents who indicated that using ChatGPT to fix grammar or rewrite text should be acknowledged. We find substantial differences in reporting views between using ChatGPT for fixing grammar and using it to rewrite text, with  22\%  of the respondents indicating that grammar correction should be reported relative to 52\% for text rewriting (a two-sided test of proportions gives p<0.0001). These perceptions were cleanly nested, as 95\% of respondents who thought grammar should be reported also thought rewriting should be reported (only 3 respondents thought that fixing grammar should be reported but not rewriting.)

\begin{figure*}[h]
\centering
    \begin{subfigure}[b]{.33\textwidth}
    \centering
    \includegraphics[width=2in]{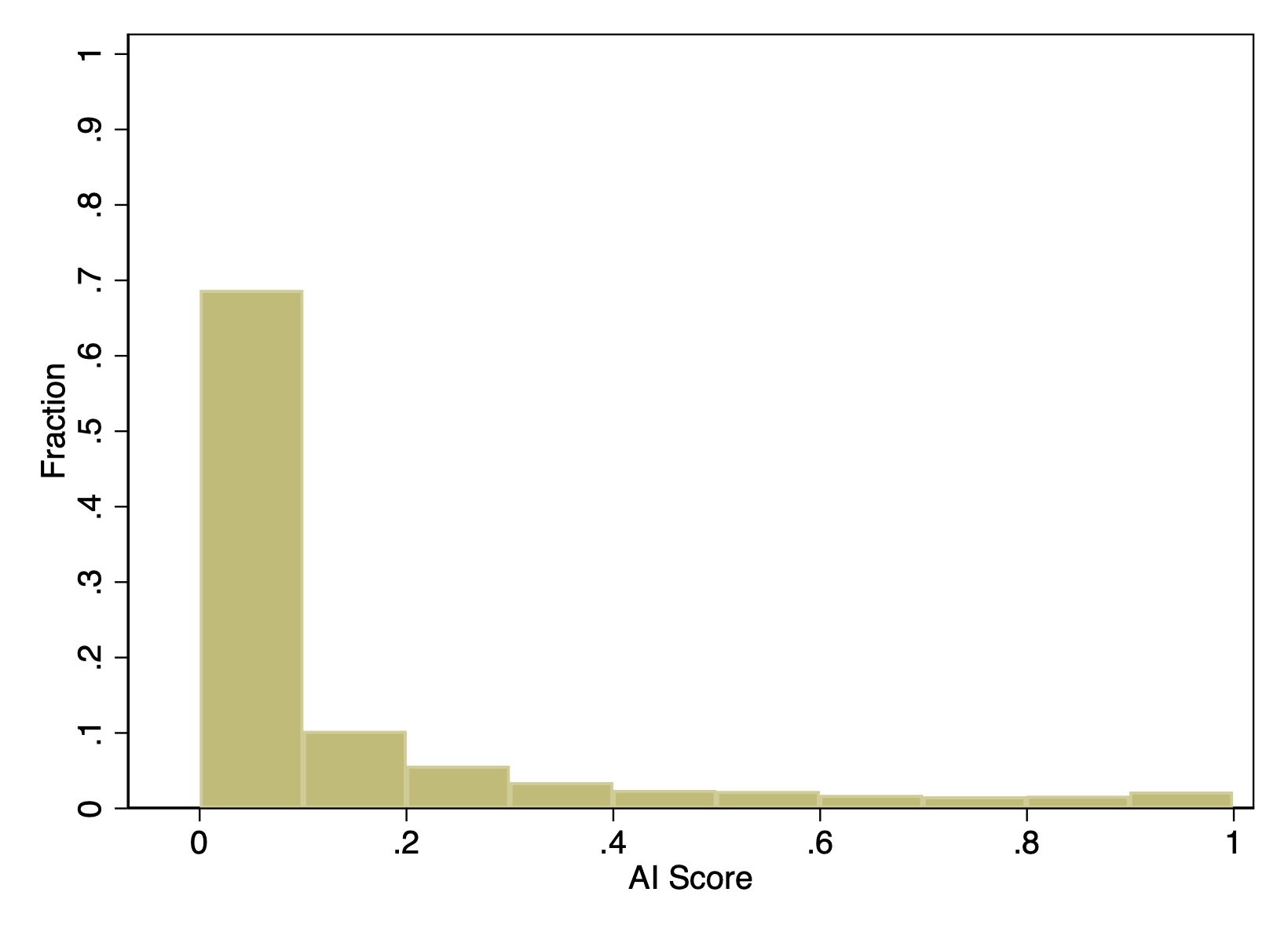}
    \caption{Original abstracts.}
    \end{subfigure}%
    \begin{subfigure}[b]{.33\textwidth}
    \centering
    \includegraphics[width=2in]{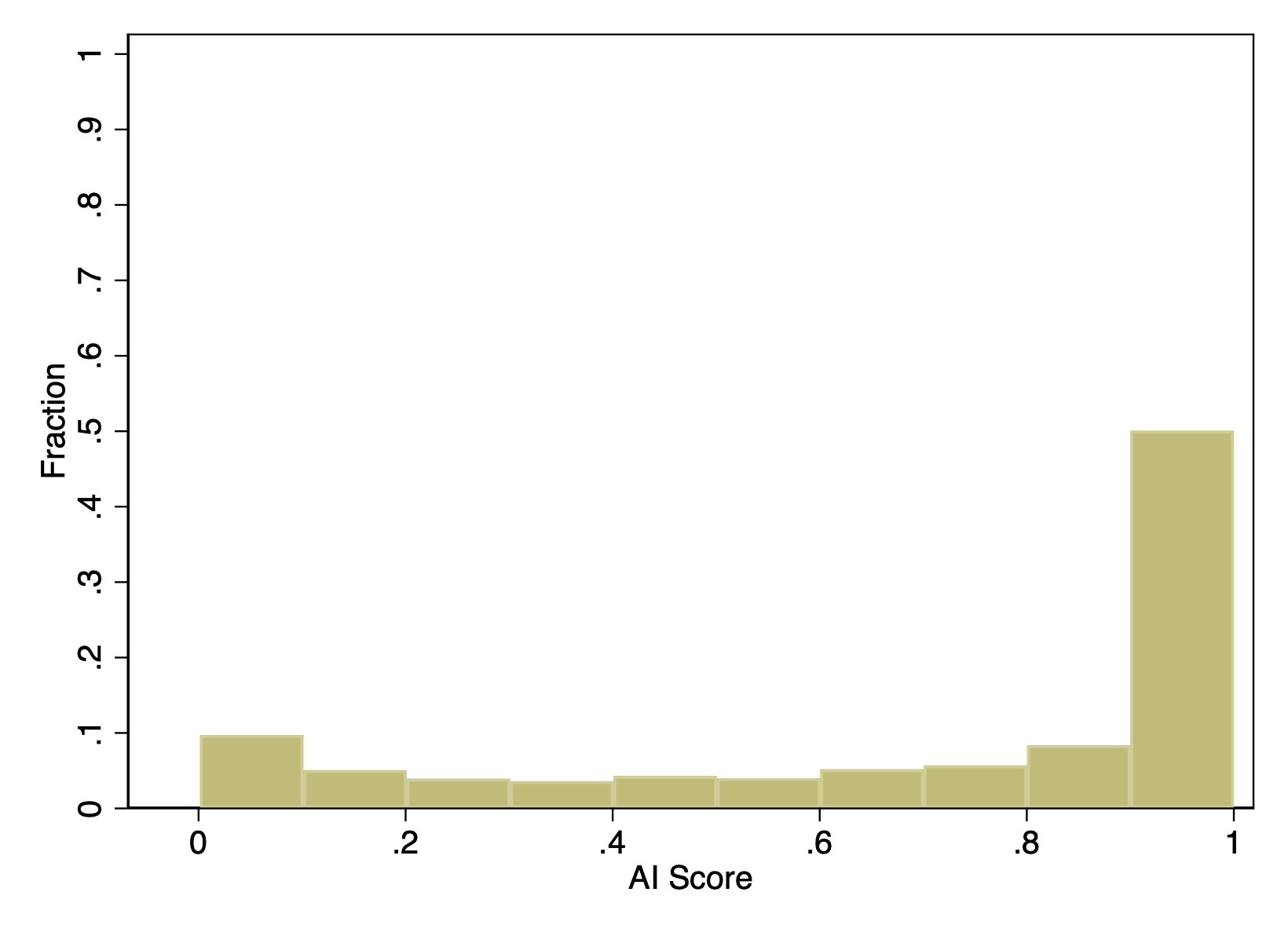}
    \caption{Grammar 1 abstracts.}
    \end{subfigure}
      \begin{subfigure}[b]{.33\textwidth}
    \centering
    \includegraphics[width=2in]{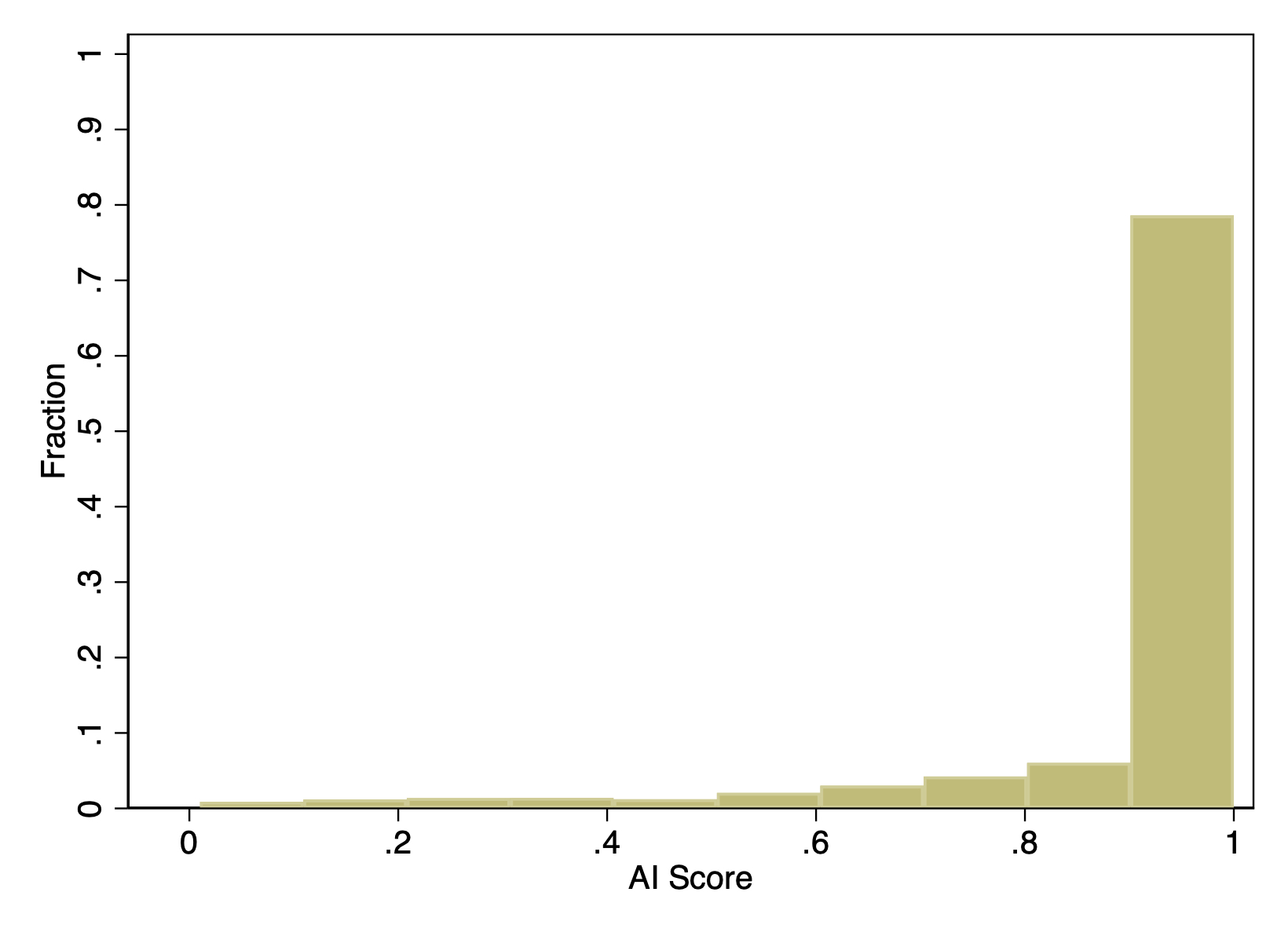}
    \caption{Rewrite 1 abstracts.}
    \end{subfigure}
    \caption{The distribution of AI scores for the original abstracts (a), abstracts revised using the Grammar 1 prompt (b), and abstracts revised using the Rewrite 1 prompt (c).}
    \label{fig:main4}
\end{figure*}

On aggregate, survey respondents viewed these types of AI use differently, but did the AI detector we study treat them differently? Figure~\ref{fig:main4} shows that the distribution of AI scores is skewed more to the right for abstracts revised using the Rewrite 1 prompt than for those revised using the Grammar 1 prompt. However, both Grammar 1 and Rewrite 1 produce a large number of high AI scores, and if we look again in Figure~\ref{fig:main1}, the 75th percentile values of the AI scores for the abstracts produced by these two prompts are both near the maximal value. It is worth noting that while the detector gave abstracts revised using both prompts high scores, gave much lower scores to the original abstracts. Thus, it had a high degree of accuracy in separating manuscripts that were not revised by ChatGPT from those that were revised in some way.

However, this analysis does not indicate how Grammar 1 and Rewrite 1 compare for a given abstract. It could be that abstracts always had higher AI scores when the Rewrite 1 prompt was used. Looking at the abstract level, 2.2\% of abstracts had the same AI score for both types of prompts and 24.2\% of abstracts had an higher AI score when the Grammar 1 prompt was used than when the Rewrite 1 prompt was used.

These results combine to produce our first main finding:
\begin{mf}
The academics we surveyed were less likely to think that using AI to fix the grammar in manuscripts should be reported than using AI to rewrite manuscripts, but detection software did not always draw this distinction, as abstracts for which GPT-3.5 was used to fix grammar were often flagged as having a high chance of being written by AI.
\end{mf}

\subsection{Heterogeneity of Perceptions}

\begin{figure*}[h]
\centering
    \begin{subfigure}[b]{.75\textwidth}
    \centering
    \includegraphics[width=1\linewidth]{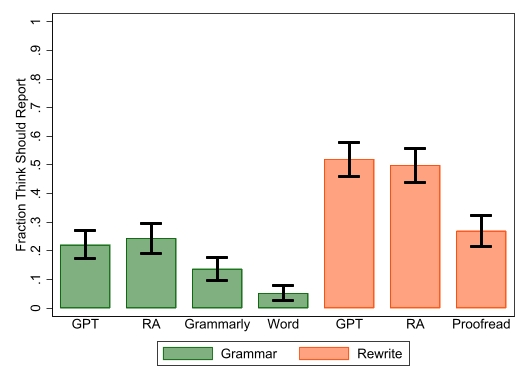}
    \end{subfigure}
    \caption{Fraction of survey respondents indicating that ChatGPT, RA, Grammarly, and Word use in fixing grammar or ChatGPT, RA, or Proofreading use in rewriting text should be reported, with 95\% confidence intervals.}
    \label{fig:main2}
\end{figure*}

When we compare our survey results for different sources of assistance, we find very similar reporting preferences between ChatGPT and the help of a research assistant (RA) for grammar correction or for rewriting text, which is illustrated in Figure~\ref{fig:main4}. More detailed summary statistics are provided in Appendix B. Comparing the difference between RA and ChatGPT for fixing grammar and rewriting text, we find no significant difference (p=0.5418 for fixing grammar and p=0.6062 for rewriting for two-sided tests of proportions).

However, respondents indicated that other tools used for fixing grammar, such as Grammarly and Word, should be acknowledged at even lower rates than ChatGPT, even though these tools might provide users with similar grammar corrections as ChatGPT. Only 14\% (5\%) of those completing the survey responded that researchers should report using Grammarly (Word) to fix grammar for academic text. A two-sided test of proportions between reporting ChatGPT and Grammarly (Word) for fixing grammar gives p=0.0100 (p<0.0001). This might indicate that researchers are unfamiliar with the differences between these tools, especially given that Grammarly uses some AI models (\cite{fitria2021grammarly}). Alternatively, respondents may be so familiar with those tools that they are less worried about their influence on academic writing compared to ChatGPT, which is relatively new.

Finally, only 27\%  of those completing the survey responded that proofreading services should be acknowledged for rewriting academic text, which is much lower than for ChatGPT (52\%)  and for RA help (49.6\%). A two-sided test of proportions comparing ChatGPT to proofreading for rewriting text gives p<0.0001. One explanation for this result could be that proofreading might be considered to be closer to fixing grammar than rewriting text (even that we explicitly stated that proofreading would be used for rewriting text). Another potential explanation could stem from an academic norm related to acknowledging this service or the fact that this service is paid. 

This leads to our second main finding:
\begin{mf}
We found little difference in preferences for reporting ChatGPT and RA help, but significant differences in reporting preferences between these sources of assistance and paid proofreading and other AI assistant tools (Grammarly and Word).
\end{mf}

\begin{figure*}[h]
\centering
    \begin{subfigure}[b]{.33\textwidth}
    \centering
    \includegraphics[width=2in]{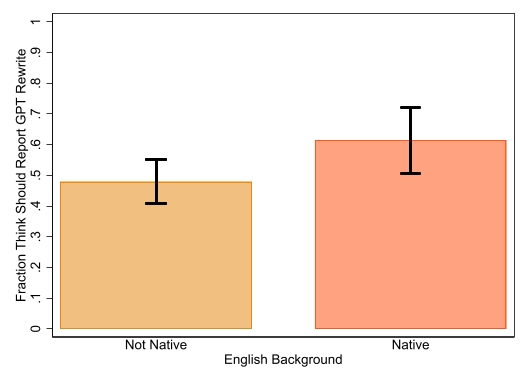}
    \end{subfigure}%
    \begin{subfigure}[b]{.33\textwidth}
    \centering
    \includegraphics[width=2in]{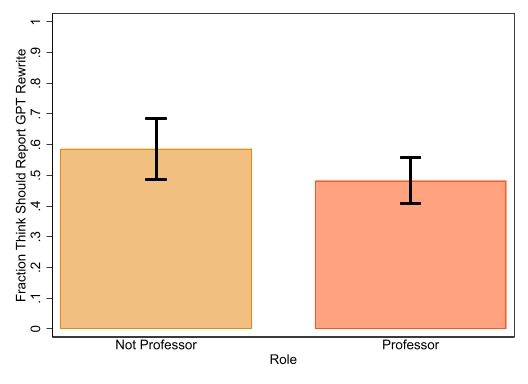}
    \end{subfigure}
      \begin{subfigure}[b]{.33\textwidth}
    \centering
    \includegraphics[width=2in]{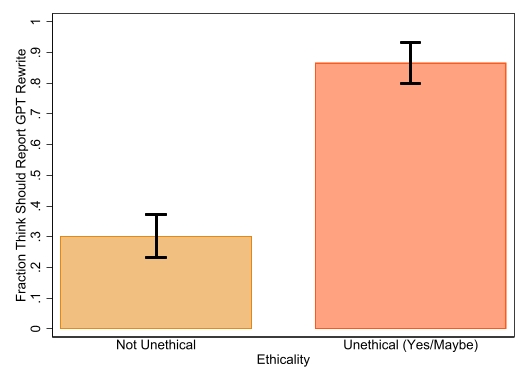}
    \end{subfigure}
    \caption{Fraction of survey respondents indicating that ChatGPT use in fixing grammar or rewriting text should be reported, with 95\% confidence intervals, by English background, academic role, and perceptions of ethics. }
    \label{fig:main3}
\end{figure*}

Next, we investigate the potential reasons why some academics we surveyed thought that using ChatGPT to rewrite text should be reported and why others did not. Figure~\ref{fig:main3} shows how reporting preferences differ by English language background (native speaker or not), academic role (professor or not), and perceptions of ethics (whether using ChatGPT to rewrite text is unethical or not). To increase statistical power, we collapse role into professor or not and pool together those who answered either ``yes'' or ``maybe'' to the survey question on the ethics of using ChatGPT to rewrite text. When we compare native English speakers to non-native ones, we find that on average native speakers are more inclined towards reporting the use of ChatGPT for rewriting text. When we compare professors (both tenured and untenured) to students and postdocs we find that, on average, postdocs and students are more likely to believe that authors should acknowledge the use of ChatGPT when rewriting text. One possible explanation is that early career researchers are more conservative because they are unfamiliar with the norms in the profession and prefer to take the safer option and report. On the other hand, they may have a better sense of the power of these tools, and hence might feel that reporting is more necessary. Finally, we find large differences in reporting preferences based on perceptions of ethics. Survey respondents who believe that it is unethical to use ChatGPT to rewrite text are almost three times more likely to believe that it should be reported. A more detailed analysis of perceptions of ethics is provided in Appendix B.

When we run an OLS regression analysis with only English language background (dummy variable Native=1) and academic role (dummy variable Professor=1) as the explanatory variables, as presented in first column of Table~\ref{tab:Reg}, we find that the coefficients are significant at the 5\% and 10\% level for Native and Professor respectively. However, when we add perceptions of ethics to the regression, the effect for Native becomes weaker and less significant. This suggests that some of this effect was due to differences in perceptions of ethics between native and non-native speakers of English.

\begin{table}[H]
\begin{center}
{
\def\sym#1{\ifmmode^{#1}\else\(^{#1}\)\fi}
\begin{tabular}{l*{2}{c}}
\hline\hline
            &\multicolumn{1}{c}{(1)}&\multicolumn{1}{c}{(2)}\\
            &\multicolumn{1}{c}{Report GPT Rewrite}&\multicolumn{1}{c}{Report GPT Rewrite}\\
\hline

Native    &       0.142\sym{**}  &      0.0851         \\
            &      (0.065)         &      (0.055)         \\
[1em]

Professor &      -0.111\sym{*}         &      -0.102\sym{*}         \\
            &     (0.063)         &     (0.052)         \\
[1em]
Unethical &                     &       0.557\sym{***}\\
            &                     &     (0.052)         \\
[1em]
Constant      &       0.547\sym{***}&       0.343\sym{***}\\
            &     (0.053)         &      (0.048)         \\
\hline
\(N\)       &         271         &         271         \\
\hline\hline
\multicolumn{3}{l}{\footnotesize Note: Standard errors in parentheses. \sym{*} \(p<0.10\), \sym{**} \(p<0.05\), \sym{***} \(p<0.01\).}\\
\end{tabular}
}
\caption{Regressions of whether using GPT to rewrite should be reported (dummy variable) onto being a native English speaker or not (dummy variable), being a professor or not (dummy variable), and in the second specification, whether using ChatGPT to rewrite text is unethical or not (dummy variable).}
\label{tab:Reg}
\end{center}
\end{table}

\begin{mf}
We found disagreements among the academics we surveyed about whether using Chat-GPT to rewrite text needs to be reported, and differences were related to perceptions of ethics, academic role, and English language background.
\end{mf}

\subsection{Detection Robustness}

\begin{figure}[H]
    \centering
    \includegraphics[width=3in]{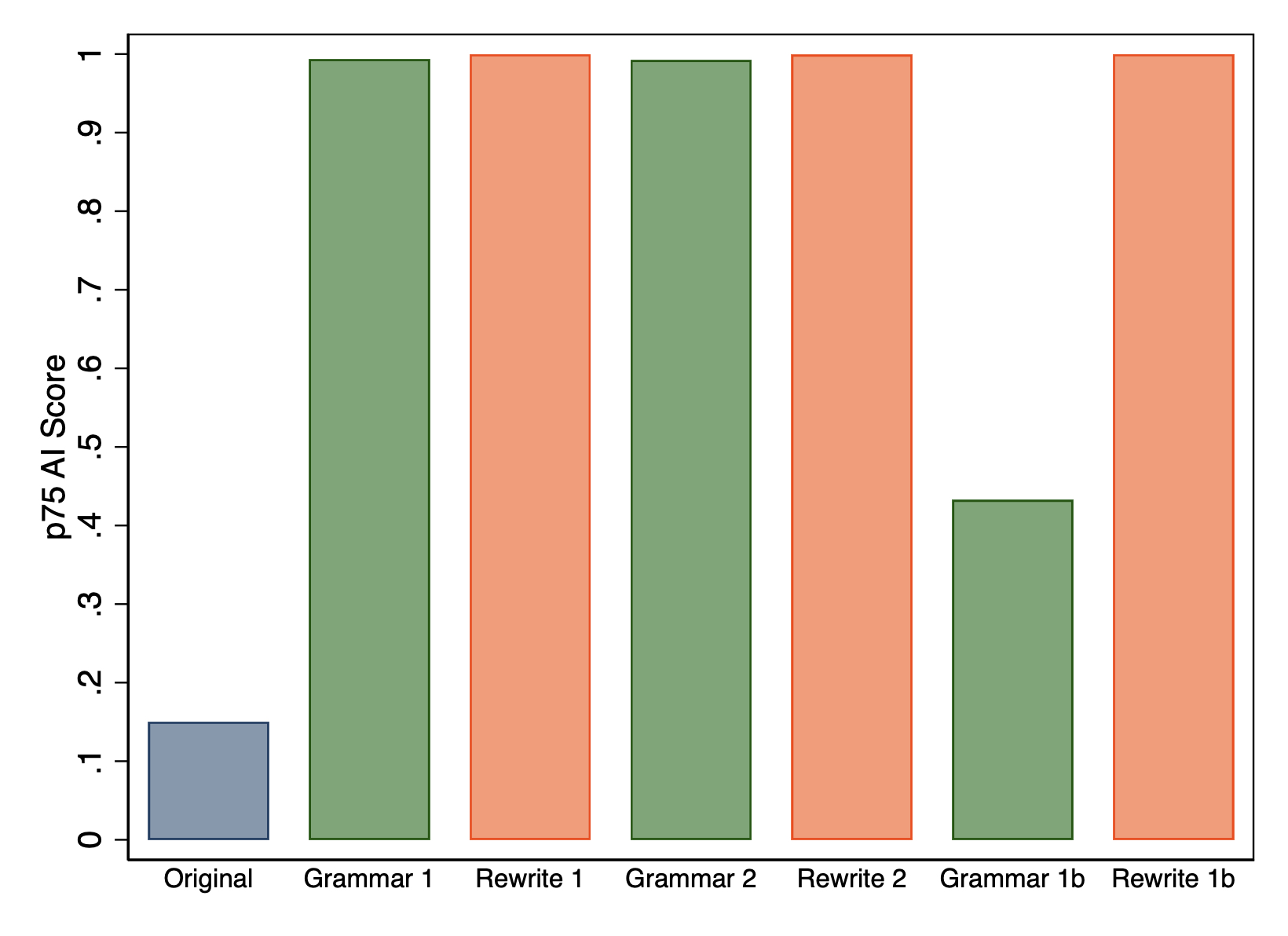}
       \caption{75th percentile of AI score for original abstracts and the versions that were revised by GPT-3.5 for all of the prompts.}
    \label{fig:AI detection scores by type 75}
\end{figure}

\begin{figure}[H]
    \centering
    \includegraphics[width=3in]{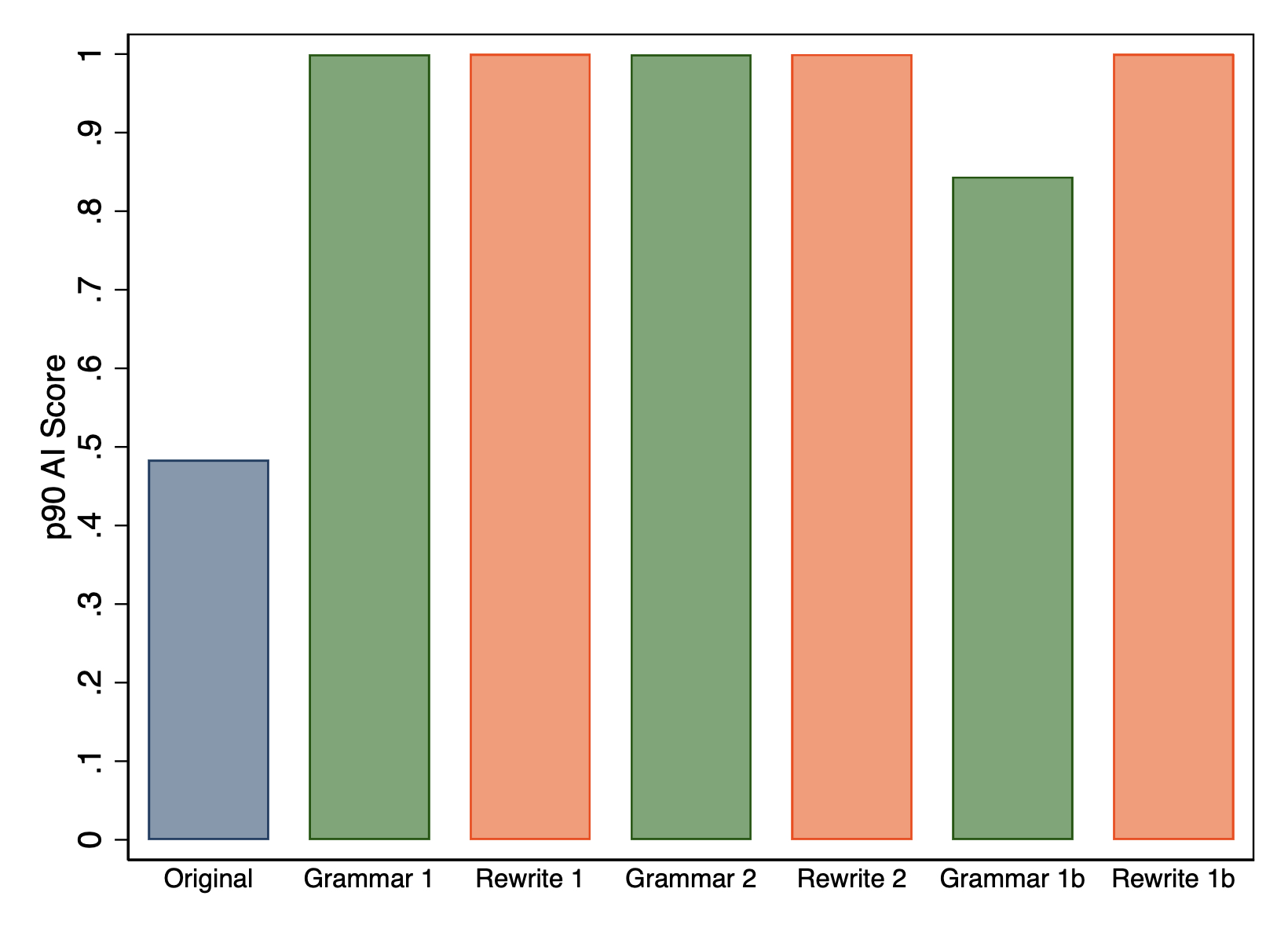}
       \caption{90th percentile of AI score for original abstracts and the versions that were revised by GPT-3.5 for all of the prompts.}
    \label{fig:AI detection scores by type 90}
\end{figure}

First, we compare the 75th percentile of AI scores across all of the prompts provided in Table~\ref{tab:table1}, and Figure~\ref{fig:AI detection scores by type 75} shows that we find no perceptible differences for Grammar 2 and Rewrite 2. However, we find that dropping the requirement that the GPT-3.5 output be only one paragraph dramatically reduces the 75th percentile value when revising grammar (Grammar 1b). However, the result is more robust if we consider the 90th percentile instead, as shown in Figure~\ref{fig:AI detection scores by type 90}.

\begin{figure}[H]
    \centering
    \includegraphics[width=3in]{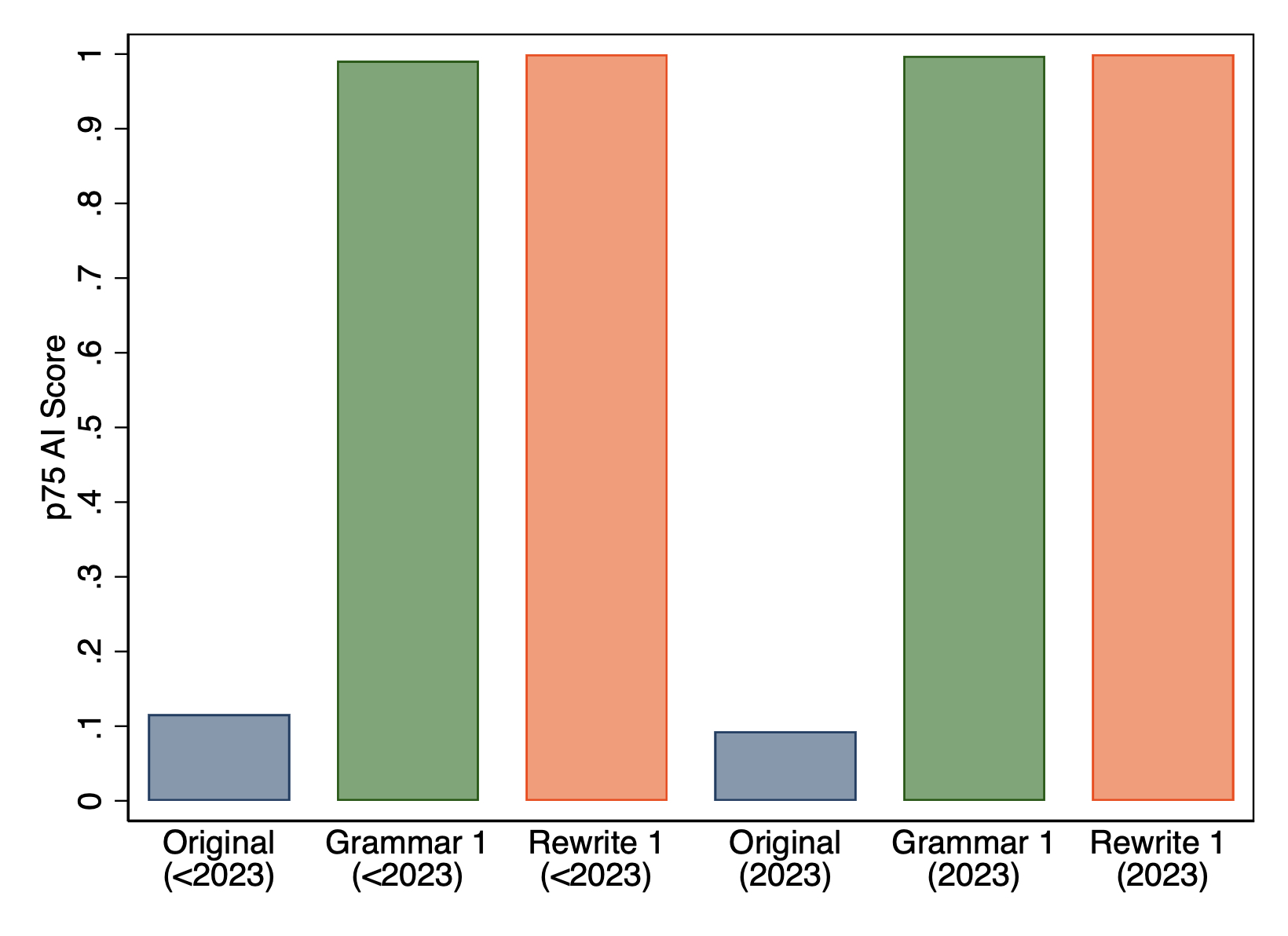}
       \caption{75th percentile of AI score for original abstracts and the versions that GPT-3.5 was used to either fix grammar or rewrite text for abstracts appearing in print before and after 2023.}
    \label{fig:years}
\end{figure}

Second, we checked if the 75th percentile values of the AI scores were different in the years before and after the launch of ChatGPT. Looking at the 75th percentile of AI scores in Figure~\ref{fig:years}, we find a very slight \textit{decrease} in the 75th percentile AI scores for original abstracts that appeared in 2023. Given this small change for original abstracts, it might not be surprisingly that we do not see much of a difference for Grammar 1 or Rewrite 1.

\begin{figure}[H]
    \centering
    \includegraphics[width=3in]{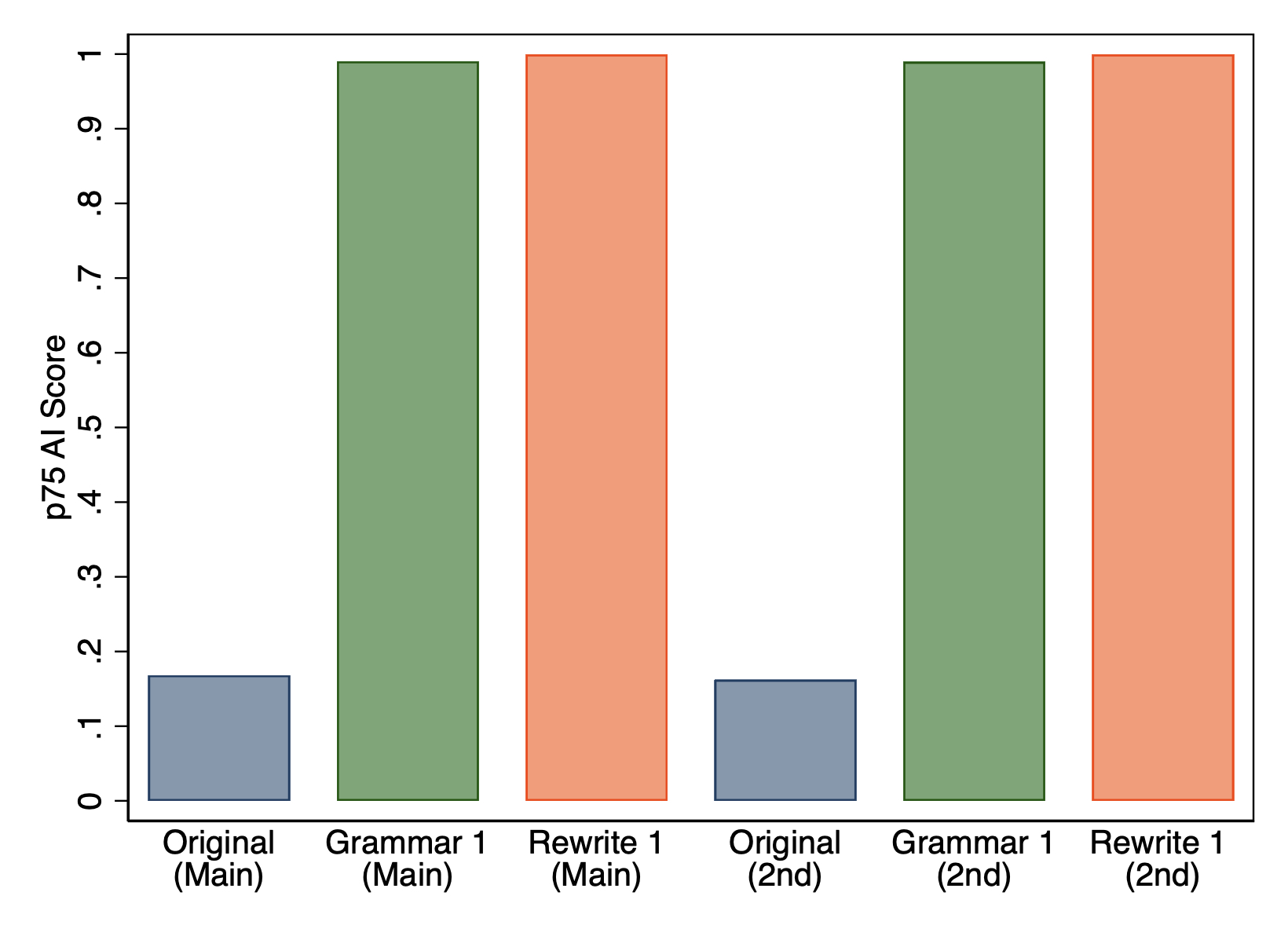}
       \caption{75th percentile of AI score for original abstracts and the versions that GPT-3.5 was used to either fix grammar or rewrite text for abstracts in our main run and in our secondary run.}
    \label{fig:second}
\end{figure}

Third, we ran 1,016 of our abstracts through ChatGPT twice to test for the variability in AI scores due to any stochasticity in ChatGPT. Figure~\ref{fig:second} shows that there is very little difference in the 75th percentile of scores.

\section{Discussion}

Our findings are a starting point for future research and suggest that several issues need to be carefully considered by the academic community. For instance, the community needs to decide which forms of assistance should be reported, whether it be ChatGPT or some other source, such as RA help or help from another AI tool. Additionally, it needs to decide which types of assistance to be reported, be it fixing grammar or something more extensive. We found more consensus that using ChatGPT to fix grammar does not need to be reported than when it comes to rewriting text. One avenue could be to disclose the actual prompts that are used to revise the paper. Along these lines, Grammarly has a new feature that allows users to acknowledge the usage of AI and the actual prompts that the users used.

In addition, our research raises the question of how journals, conferences, and associations can enforce differences in what should be reported. While the detection tool that we employed was able to detect relatively accurately whether AI was used at all, abstracts that were rewritten by GPT-3.5 were sometimes given a lower chance of being written by AI than the grammar-fixed abstracts. This also opens the question of whether using ChatGPT to fix grammar might inadvertently make more substantial changes than desired. How can researchers be sure that they use AI in the desired way? One solution could be to use ChatGPT to point out grammar errors, but to fix them manually, so that ChatGPT does not actually revise the text.

These tools and our perceptions of them will surely evolve, but the aim of this paper is to determine how they are perceived and detected in this moment in time, as it appears to be an inflection point in AI ability and in its use to revise text.

\subsection{Limitations and Future Directions}

In terms of our survey design, the use of a convenience sample may have introduced selection issues that complicate our comparisons by role and English language background. In addition, our convenience sample was largely composed of economists, and because views might differ across fields, it would be valuable to also consider what perceptions look like more generally and to illuminate differences across fields. An alternative approach that allows for an assessment of perceptions across fields is offered by \textcite{bringula2023academics}, who runs a sentiment analysis on papers written about AI use in manuscript preparation and finds that the sentiment in those papers is generally positive. Also, the use of a convenience sample did not lend itself to randomizing on the form of assistance (ChatGPT, RA assistance, etc.).

In addition, another limitation of our survey is that we do not dig deeply into the nature of ethical perceptions. Since these perceptions were such an important predictor of reporting preferences, it might be insightful to know why academics feel that using AI tools for manuscript preparation is unethical. For example, is someone harmed by their use -- such as other academics, science in general, or the authors whose material is used in training the AI -- or is there a deeper moral question at play? One way to tease apart these subtleties would be to have academics evaluate a number of detailed vignettes. It might also be interesting to determine the role of payment in ethical considerations, as ChatGPT, RAs, and proofreaders sometimes require payment and sometimes do not.

In terms of our detection design, we just considered published papers, at a top journal, and for the field of management. To understand whether these results hold more generally, it would be necessary to look at papers published in other fields, perhaps using a service like Scopus, and to look at working papers, perhaps sourced from SSRN or arXiv. In addition, it might be valuable to consider full papers instead of just abstracts, or as a middle case between abstracts and full papers, introductions might also be useful to examine.

Additionally, our results are limited to one AI detection service, so we do not know if they extend to other services, such as GPTZero. Also, it might be of interest to see if other AI-based revision services, such as Grammarly, are flagged by AI detectors too. Along these lines, it might be insightful to consider other forms of writing assistance that might be taken besides fixing grammar and rewriting text, or specific forms of rewriting text. One important dimension could be whether use of chat-based and completion-based AI tools lead to different detection rates. 

Finally, and related to the last point, it would be valuable to consider a range of different prompts, especially given the sensitivity that we found to relatively small changes in prompt language. One systematic way to choose the prompts could be to hold focus groups of researchers or ChatGPT users or by having a way for researchers to vote on the prompts that are tested. Another dimension of investigation could be to see if our results are sensitive to other important features of the GPT API inputs, such as increasing the model to GPT-4, increasing the temperature to increase the hallucination rate, or by using other system prompts. Given the black box nature of LLMs, a robust empirical analysis is needed to inform policymakers at associations, journals, and conferences of the link between how AI is used and how it is flagged by detection software.

\pagebreak
\printbibliography

\newpage
\appendix
\clearpage
\begin{center}
\textbf{\Large{Appendix for \textit{Perceptions and Detection of AI Use in Manuscript Preparation for Academic Journals}}}    
\end{center}\section{Survey Screenshots}
\begin{figure}[H]
    \centering
    \includegraphics[width=6in]{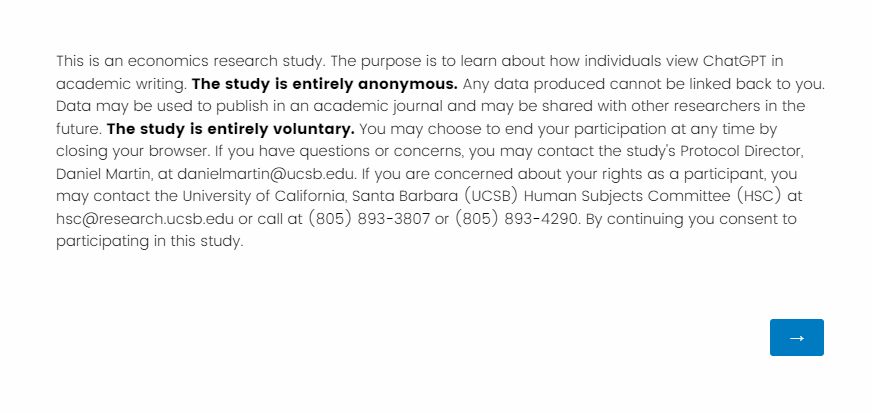}
     \caption{Consent form}
    \label{fig:Survey_introl}
\end{figure}
    \begin{figure}[H]
    \centering
    \includegraphics[width=5in]{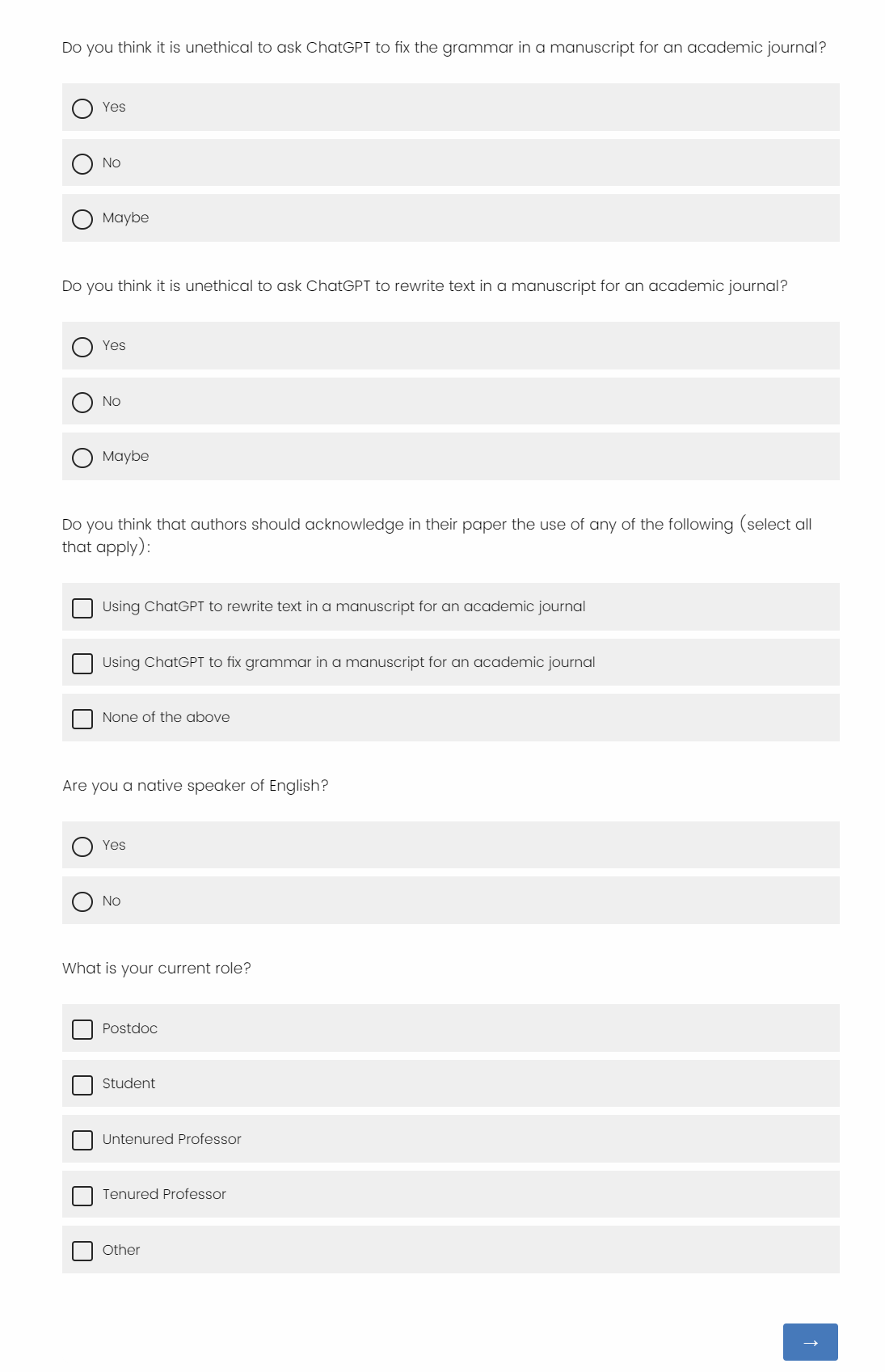}
        \caption{Survey questions (page 1)}
         \label{fig:part1}
       \end{figure}
 \begin{figure}[H]
    \centering
    \includegraphics[width=5in]{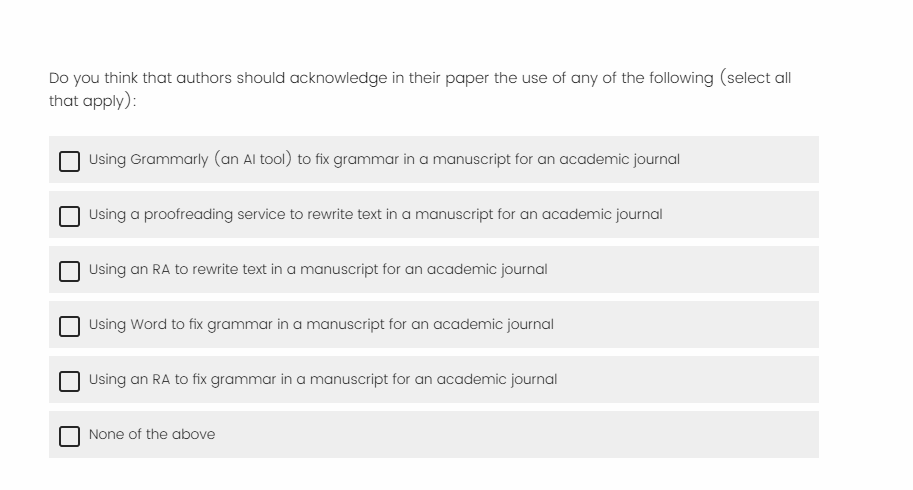}
       \caption{Survey questions (part 2)}
    \label{fig:part2}
\end{figure}

\section{Additional Tables}

Table~\ref{tab:acknowledge} provides summary statistics for whether authors should acknowledge different forms of assistance. The top panel provides the average responses broken out by being a native English speaker or not, and the bottom panel by being a professor or not.

\begin{table}[H]
\begin{center}
  \renewcommand{\arraystretch}{0.6} 
  \setlength{\tabcolsep}{2pt} 
\begin{tabular}{@{}lcccccccc@{}}
                & \multicolumn{4}{c}{Grammar} &  \multicolumn{3}{c}{Rewrite} &Respondents \\ 
                   &ChatGPT & RA & Word &  Grammarly  & ChatGPT& RA & Proofreading & \\ 
                   &(mean)  & (mean)  & (mean)  &  (mean)   & (mean) & (mean)  & (mean)  & (N)\\ 
\midrule
\textbf{Native} &&&&\\[1ex]
Yes  &0.27&0.27&0.05&0.17&0.61&0.53&0.28&83\\[1ex]
No  &0.2&0.23&0.05&0.12&0.48&0.48&0.27&188\\[1ex]
\midrule
\textbf{Total}  &\textbf{0.22}&\textbf{0.24}&\textbf{0.05}&\textbf{0.14}&\textbf{0.52}&\textbf{0.5}&\textbf{0.27}&\textbf{271}\\[1ex]
\midrule
\textbf{Role} &&&&\\[1ex]
Professor  &0.2&0.21&0.04&0.1&0.48&0.47&0.2&172\\[1ex]
Student  &&&&&&&&\\
or Postdoc &0.25&0.3&0.07&0.2&0.59&0.56&0.38&99\\[1ex]
\midrule
\textbf{Total}  &\textbf{0.22}&\textbf{0.24}&\textbf{0.05}&\textbf{0.14}&\textbf{0.52}&\textbf{0.5}&\textbf{0.27}&\textbf{271}\\[1ex]
\end{tabular}
\caption{The survey averages responses to whether authors should acknowledge using ChatGPT, RA, Grammarly, and Word for grammar correction, as well as ChatGPT, RA, and proofreading for text rewriting based on the respondent's role and whether the respondent is a native English speaker.}\label{tab:acknowledge}
\end{center}
\end{table}

Table~\ref{tab:Unethical} shows the results of asking if it is unethical to use ChatGPT to correct the grammar or rewrite the text in a manuscript for an academic journal. For fixing grammar, there is a consensus that it is ethical to use ChatGPT, with 94\%  of the researchers replying that it is not unethical. However, for rewriting, this becomes more open questions with only 61\% respondents thinking that is not unethical. We allow respondents to answer ``maybe'' to these questions to identify which researchers are uncertain about this question. However, in our analysis, we categorize respondents into those who believe it is ethical and those who either answered that it is unethical or indicate that it is ``maybe'' unethical.

\begin{table}[H]
\centering
  \renewcommand{\arraystretch}{0.7} 
  \setlength{\tabcolsep}{2pt} 
\begin{tabular}{@{}lccc@{}}
                   &Unethical Grammar&Unethical Rewrite& Respondents\\ 
                   &(mean)  &(mean) & (N)\\ 
\midrule
\textbf{Native} &&&\\[1ex]
Yes  &0.07&0.46&83\\[1ex]
No   &0.05&0.36&188\\[1ex]
\midrule
\textbf{Total} &\textbf{0.06}&\textbf{0.39}&\textbf{271}\\[1ex]
\midrule
\textbf{Role} &&&\\[1ex]
Professor &0.06&0.38&172\\[1ex]
Student or Postdoc &0.05&0.39&99\\[1ex]
\midrule
\textbf{Total} &\textbf{0.06}&\textbf{0.39}&\textbf{271}\\[1ex]
\end{tabular}
\caption{The survey averages responses to the questions: ``Do you think it is unethical to ask ChatGPT to fix grammar (rewrite text) in a manuscript for an academic journal?'' based on one's role and whether they are a native English speaker.}\label{tab:Unethical}
\end{table}

When we compare native English speakers to non-native ones, we find that there is not a significant difference in perceptions of ethics around fixing grammar: 7\% compared to 5\% (two-sample test of proportions yields p=0.211). However, when it comes to rewriting text, 46\% of native English speakers think that it is unethical to use ChatGPT, compared to 36\% of non-native speakers. A two-sample test of proportions yields p = 0.121 on a two-tailed test or p = 0.060 on the one-tailed. When we compare professors to students and postdocs, we do not find any significant difference for either fixing grammar and rewriting text (for grammar p=0.76 and for rewrite p=0.977). 

Finally, Table~\ref{tab:Acknowledgeunethical} shows the perceptions of respondents regarding whether one should report the use of the ChatGPT for assistance in manuscript preparation. This includes those who consider it ethical to use it and those who believe it to be unethical or are unsure (answered ``maybe''). A significant difference exists among these groups. When researchers perceive it as ethical to use ChatGPT, only 21\% (30\%) believe that reporting is necessary when using it for grammar correction (rewriting). In contrast, researchers who view it as unethical or are uncertain about its ethics have a higher percentage, with 47\% (87\%) believing that reporting is necessary. This difference among the groups is significant. A two-sample test of proportions yields p=0.017 (0.0001).

\begin{table}[H]
\begin{center}
  \renewcommand{\arraystretch}{0.7} 
  \setlength{\tabcolsep}{2pt} 
\begin{tabular}{@{}lcc@{}}
                   &Acknowledge & Respondents\\ 
                   &(mean) & N\\ 
\midrule
\textbf{Grammar} &&\\[1ex]
Unethical (Yes or Maybe) &0.47&15\\[1ex]
Unethical (No)  &0.21&256\\[1ex]
\midrule
\textbf{Total} &\textbf{0.22}&\textbf{271}\\[1ex]
\midrule
\textbf{Rewrite} &&\\[1ex]
Unethical (Yes or Maybe) &0.87&105\\[1ex]
Unethical (No) &0.3&166\\[1ex]
\midrule
\textbf{Total} &\textbf{0.52}&\textbf{271}\\[1ex]
\midrule
\end{tabular}
\caption{The average perception of researchers regarding whether authors should report using the ChatGPT for assistance in manuscript preparation is grouped by whether the respondent thinks it is unethical to use it to fix grammar or rewrite the academic text.}\label{tab:Acknowledgeunethical}
\end{center}
\end{table}

\pagebreak
\begin{landscape}
\begin{table}[H]
\centering\small
\begin{tabular}{lll}
Publisher & Disclosure & Link\\ 
\hline
Science &Mandatory& https://www.science.org/content/page/science-journals-editorial-policies\\

Elsevier & Mandatory& https://www.elsevier.com/about/policies-and-standards/publishing-ethics\\

Springer Nature & Voluntary & https://www.nature.com/nature/editorial-policies\\

Taylor \& Francis & Mandatory & https://newsroom.taylorandfrancisgroup.com/taylor-francis-clarifies-the-responsible-use-of-ai-tools-in-academic-content-creation/\\

JAMA Network & Voluntary& https://jamanetwork.com/journals/jama/fullarticle/2801170\\

American Chemistry Society (ChemArxiv) & Mandatory& https://axial.acs.org/publishing/new-chemrxiv-policy-on-the-use-of-ai-tools\\

International Committee of Medical Journal & Mandatory& https://www.icmje.org/icmje-recommendations.pdf\\

World Association of Medical Editors & Mandatory& https://wame.org/page3.php?id=106\\

International Conference on Machine Learning & Mandatory& https://www.acm.org/publications/policies/new-acm-policy-on-authorship\\

\hline
\end{tabular}
\caption{AI disclosure policy for several major publishers as of November 18, 2023.}\label{tab:Publishers}
\end{table}
\end{landscape}

\end{document}